\documentclass[onecolumn,showpacs,fleqn,nobibnotes]{revtex4}

\usepackage{amsmath}
\usepackage{graphicx}
\usepackage{float}
\usepackage{subfigure}

\def\lsim{\raise0.3ex\hbox{$<$\kern-0.75em\raise-1.1ex\hbox{$\sim$}}}
\def\gsim{\raise0.3ex\hbox{$>$\kern-0.75em\raise-1.1ex\hbox{$\sim$}}}
\def\pom{{I\!\!P}}
\newcommand{\beq}{\begin{equation}}
\newcommand{\eeq}{\end{equation}}

\begin{document}

\title{Heavy quark production in $\gamma \pom$ interactions at hadronic colliders}
\author{V.P. Gon\c{c}alves
and M.M. Machado}

\affiliation{Instituto de F\'{\i}sica e Matem\'atica, Universidade Federal de
Pelotas\\
Caixa Postal 354, CEP 96010-900, Pelotas, RS, Brazil.}

\begin{abstract}
The diffractive heavy quark cross sections are estimated considering photon-Pomeron ($\gamma \pom$) interactions in hadron - hadron at RHIC, Tevatron, and CERN LHC energies. We assume the validity of the hard diffractive factorization and calculate the charm and bottom total cross sections and rapidity distributions using the diffractive parton distribution functions of the Pomeron obtained by the H1 Collaboration at DESY-HERA.  Moreover, we compare our predictions with those obtained using the dipole model. We verify that this process is a good test of the different mechanisms for diffractive  heavy quark production at hadron - hadron colliders. 
\end{abstract}
\maketitle
\section{Introduction}

A long-standing puzzle in the particle physics is the nature of the Pomeron ($\pom$). This object, with the quantum numbers of the vacuum, was introduced phenomenologically in the Regge theory as a simple moving pole in the complex angular momentum plane, to describe the high-energy behavior of the total and elastic cross-sections of the hadronic reactions \cite{collins}. Within the framework of the perturbative Quantum Chromodynamics (pQCD), the Pomeron is associated with the resummation of leading logarithms in $s$ (center of mass energy squared) and at lowest order is described by the two-gluon exchange \cite{PREDAZZI}. Due to its zero color charge the Pomeron is associated with diffractive events, characterized by the presence of large rapidity gaps in the hadronic final state. The diffractive processes have attracted much attention as a way of amplifying the physics programme at hadronic colliders, including searching for New Physics (For recent reviews see, e.g. Refs. \cite{scho,forshaw}). The investigation of these reactions at high energies gives important information about  the structure of hadrons and their interaction mechanisms. In particular, hard diffractive processes allow the study of the interplay of small- and large-distance dynamics within Quantum Chromodynamics (QCD). 

The diffractive processes can be classified as inclusive or exclusive events (See e.g. \cite{forshaw}).
In exclusive events, empty regions  in pseudo-rapidity, called rapidity gaps, separate the intact very forward hadron from the central massive object. Exclusivity means that nothing else is produced except the leading hadrons and the central object.
The inclusive processes also exhibit rapidity gaps. However,  they contain soft particles accompanying the production of a hard diffractive object, with the rapidity gaps becoming,  in general, smaller than in the exclusive case. It is important to emphasize that at the LHC, the only possibility to  detect double diffractive events is by tagging the intact hadrons in the final state, because of the pile up of events in each bunching crossing. It implies the key element to measure diffractive events at the LHC will be tagging  the  forward scattered incoming hadrons. 



A good testing ground for diffractive physics is the heavy quark production. In the last years, several authors have discussed this process considering different approaches. In particular, in Refs. \cite{mag_hqdif,MMM1,MMM2} the total cross sections for Pomeron - hadron (inclusive single diffraction)  and Pomeron - Pomeron (inclusive central diffraction)  heavy quark production were calculated assuming the validity of the diffractive factorization formalism and that Pomeron has a partonic structure. The basic idea is that  the hard scattering resolves the quark and gluon content in the Pomeron \cite{IS} and  it can be obtained analysing the experimental data from diffractive deep inelastic scattering (DDIS) at HERA, providing us with the diffractive distributions of singlet quarks and gluons in the Pomeron \cite{H1diff}. This model is usually denoted resolved Pomeron model. On the other hand, in Ref. \cite{kope} the diffractive heavy quark production was estimated within the light-cone dipole approach, which is able to describe inclusive and diffractive processes in a unified way, with the  key ingredient being the dipole-target cross section determined from HERA $F_2$ data. One of the main conclusions of that paper is the breakdown of factorization in diffractive hadronic collisions, which put in check the calculations presented in Refs. \cite{mag_hqdif,MMM1,MMM2}. However, as both models describe the experimental data for single diffractive charm and bottom production, the correct approach to describe the inclusive diffraction process remains an open question. Finally, more recently, in Refs. \cite{antoni_prd82,antoni_plb}, the exclusive double diffractive production of heavy quarks was estimated considering an improved version of the Durham model  \cite{kmr_first,kmr} (For a review see, e.g. \cite{kmr_review})  and obtained that the cross section for this mechanism is much higher than the cross section for inclusive double-Pomeron interactions calculated in \cite{mag_hqdif}.  Consequently, the present scenario for diffractive heavy quark production is unclear, motivating the study of  alternative processes which allow to constrain the correct description of the Pomeron.

An alternative is the study of diffractive heavy quark production in hadronic interactions mediated by photons, usually denoted coherent or ultraperipheral collisions in the literature (For reviews see Ref. \cite{upc}).  
Basically, in hadron - hadron colliders, the relativistic hadrons  give rise to strong electromagnetic fields. The photon stemming from the electromagnetic field of one of the two colliding hadrons can interact with one photon of
the other hadron (two - photon process) or can  interact directly with the other hadron (photon - hadron
process). In the latter case, it is possible to study photon - parton (inclusive) and photon - Pomeron (diffractive) processes.
Consequently, coherent processes allow us to study the inclusive and diffractive heavy quark photoproduction. In principle, the experimental signature of these processes is distinct and it can easily be separated. While in  the inclusive heavy quark photon-hadron production  we expect only one rapidity gap (associated to the photon exchange) and the dissociation of one of the incident hadrons, in diffractive  heavy quark photoproduction,  we expect the presence of two rapidity gaps and hadrons intact in the final state, similarly to exclusive double diffractive production. An important aspect to be emphasized is that in contrast to heavy quark production in hadron - Pomeron and Pomeron - Pomeron interactions, which have its predictions strongly affected by the soft absorption corrections, in photon - Pomeron interactions these corrections are expected to be very small.
In Refs. \cite{vicmag_perhq,vicmag_inc,vicmag_hqdif,vicmag_ane} heavy quark photoproduction was studied in detail, considering the dipole approach, and it was advocated that the cross sections for inclusive and diffractive interactions   may be substantial (See also Refs. \cite{vicbert,bert}). In these diffractive calculations the two-gluon perturbative model for the Pomeron was assumed. 
As the correct description of the Pomeron still is an open question, in this paper we will revisit this process and calculate the diffractive heavy photoproduction cross section  assuming that the Pomeron has a partonic structure (resolved Pomeron model) and compare the predictions of both approaches for the total cross sections and rapidity distributions for hadronic collisions  at RHIC, Tevatron and LHC energies. Our goal is twofold: (a) verify if the diffractive heavy quark photoproduction can be useful to discriminate between the resolved and perturbative models for the Pomeron and (b) compare our predictions with those obtained in Refs. \cite{mag_hqdif,kope,antoni_plb}. 

The paper is organized as follows. In the next section we present a brief review of coherent interactions in hadronic colliders. In Section \ref{sec:pomeron} we discuss the resolved Pomeron model and in Section \ref{sec:results} we present our results. Finally, in Section \ref{sec:conc} we summarize our main conclusions.

\section{Coherent interactions in hadronic colliders}

Lets consider the hadron-hadron interaction at large impact parameter ($b > R_{h_1} + R_{h_2}$) and at ultra relativistic energies. In this regime we expect the electromagnetic interaction to be dominant.
In  heavy ion colliders, the heavy nuclei give rise to strong electromagnetic fields due to the coherent action of all protons in the nucleus, which can interact with each other. In a similar way, it also occurs when considering ultra relativistic  protons in $pp(\bar{p})$ colliders.
The photon stemming from the electromagnetic field
of one of the two colliding hadrons can interact with one photon of
the other hadron (two-photon process) or can interact directly with the other hadron (photon-hadron
process). One has that  the total
cross section for a given process can be factorized in terms of the equivalent flux of photons of the hadron projectile and  the photon-photon or photon-target production cross section \cite{upc}. In general, the cross sections for $\gamma h$ interactions are at least two order of magnitude larger than for $\gamma \gamma$ interactions (See e.g. Ref. \cite{vicmag_perhq}). In what follows our main focus shall be in photon - hadron processes.
The cross section for the diffractive  photoproduction of a final state $Q\bar{Q}$ (= $c\bar{c}$ or $b\bar{b}$) in a coherent  hadron-hadron collision is  given by,
\begin{eqnarray}
\sigma (h_1 h_2 \rightarrow h_1 \otimes Q \bar{Q} X \otimes h_2) = \int \limits_{\omega_{min}}^{\infty} d\omega  \,\frac{dN_{\gamma/h_1}(\omega)}{d\omega}\,\sigma_{\gamma h_2 \rightarrow Q \bar{Q} X h_2} (W_{\gamma h_2}^2) + \int \limits_{\omega_{min}}^{\infty} d\omega  \,\frac{dN_{\gamma/h_2}(\omega)}{d\omega}\,\sigma_{\gamma h_1 \rightarrow Q \bar{Q} X h_1} (W_{\gamma h_1}^2),
\label{sighh}
\end{eqnarray}
where $\otimes$ represents a rapidity gap in the final state, $\omega$ is the photon energy,   $\omega_{min}=M_{X}^2/4\gamma_L m_p$,  $\gamma_L$ is the Lorentz boost  of a single beam, $\frac{dN_{\gamma}}{d\omega}$ is the equivalent photon flux, $W_{\gamma h}^2=2\,\omega \sqrt{S_{\mathrm{NN}}}$  and $\sqrt{S_{\mathrm{NN}}}$ is  the c.m.s energy of the
hadron-hadron system.
Considering the requirement that  photoproduction
is not accompanied by hadronic interaction (ultra-peripheral
collision) an analytic approximation for the equivalent photon flux of a nuclei can be calculated, which is given by \cite{upc}
\begin{eqnarray}
\frac{dN_{\gamma/A}\,(\omega)}{d\omega}= \frac{2\,Z^2\alpha_{em}}{\pi\,\omega}\, \left[\bar{\eta}\,K_0\,(\bar{\eta})\, K_1\,(\bar{\eta})+ \frac{\bar{\eta}^2}{2}\,{\cal{U}}(\bar{\eta}) \right]\,
\label{fluxint}
\end{eqnarray}
where  $\eta
= \omega b/\gamma_L$; $K_0(\eta)$ and  $K_1(\eta)$ are the
modified Bessel functions.
Moreover, $\bar{\eta}=\omega\,(R_{h_1} + R_{h_2})/\gamma_L$ and  ${\cal{U}}(\bar{\eta}) = K_1^2\,(\bar{\eta})-  K_0^2\,(\bar{\eta})$.
The Eq. (\ref{fluxint}) will be used in our calculations of heavy quark  production in $pA$ collisions. On the other hand, for   proton-proton collisions, we assume that the  photon spectrum of a relativistic proton is given by  \cite{Dress},
\begin{eqnarray}
\frac{dN_{\gamma/p}(\omega)}{d\omega} =  \frac{\alpha_{\mathrm{em}}}{2 \pi\, \omega} \left[ 1 + \left(1 -
\frac{2\,\omega}{\sqrt{S_{NN}}}\right)^2 \right] 
\left( \ln{\Omega} - \frac{11}{6} + \frac{3}{\Omega}  - \frac{3}{2 \,\Omega^2} + \frac{1}{3 \,\Omega^3} \right) \,,
\label{eq:photon_spectrum}
\end{eqnarray}
with the notation $\Omega = 1 + [\,(0.71 \,\mathrm{GeV}^2)/Q_{\mathrm{min}}^2\,]$ and $Q_{\mathrm{min}}^2= \omega^2/[\,\gamma_L^2 \,(1-2\,\omega /\sqrt{S_{NN}})\,] \approx (\omega/
\gamma_L)^2$.

Some comments are in order here. Firstly,  the coherence condition limits the photon virtuality to very low values, which implies that for most purposes, they can be considered as real. Moreover, if we consider
$pp/pPb$ collisions at LHC, the Lorentz factor  is
$\gamma_L = 7455/4690 $, giving the maximum c.m.s. $\gamma N$ energy
$W_{\gamma p} \approx 8390/1500$ GeV. Therefore, while studies of photoproduction at HERA were limited to photon-proton center of mass energies of about 200 GeV, photon-hadron interactions at  LHC can reach one order of magnitude higher on energy.    Secondly, in $pA$ interactions one have that due to the asymmetry in the collision, with the ion being likely the photon emitter, the photon direction is known, which will implicate an asymmetry in the rapidity distribution (see below).
Finally, in the resolved Pomeron model, in which the Pomeron has a partonic structure, the state $X$ in the final state is associated to the hadronization of the Pomeron remnants. In contrast, in the perturbative Pomeron model analysed in  \cite{vicmag_hqdif,vicmag_ane}, the final state $X$ is not  present.

\section{Heavy Quark Photoproduction in the Resolved Pomeron Model}
\label{sec:pomeron}

In the resolved Pomeron model  the diffractive cross sections are given in terms of parton distributions in the Pomeron and a Regge parametrization of the flux factor describing the Pomeron emission. The  parton distributions has evolution given by the DGLAP evolution equations and   are determined from events with a rapidity gap or a intact proton, mainly at HERA.
The basic idea to describe the diffractive heavy quark photoproduction in the resolved Pomeron model is that the cross section can be described similarly to the inclusive heavy quark photoproduction, with a diffractive gluon distribution $g^D (x,\mu^2)$ replacing the standard inclusive gluon distribution (See e.g. Refs. \cite{martinwus,vicmag_cha}). Explicitly one have
\begin{eqnarray}
\sigma(\gamma h \rightarrow Q\bar{Q}X h) (W_{\gamma h}) =    \int_{x_{min}}^1 dx \sigma^{\gamma g \rightarrow Q\bar{Q}} (W_{\gamma g}) \,g^D (x,\mu^2) 
\end{eqnarray}
with $W_{\gamma h}$ the center-of-mass energy of the photon-hadron system, $x$ the momentum fraction carried by the gluon, $x_{min} = 4 m_Q^2/W_{\gamma h}^2$ and $m_Q$ is the mass of the heavy quark. Moreover, the cross section for the photon-gluon fusion is given at leading order by \cite{hqphoto1,hqphoto2,hqphoto3}
\begin{eqnarray}
\sigma^{\gamma g \rightarrow Q\bar{Q}}(W_{\gamma g}) = \frac{2 \pi \alpha_{em} \alpha_s(\mu^2) e_Q^2}{W_{\gamma g}^2} \left[ \left( 1 + \beta - \frac{1}{2} \beta^2 \right) \, \ln \left( \frac{1+\sqrt{1-\beta}}{1-\sqrt{1-\beta}}\right) - (1+\beta)\sqrt{1-\beta}\right]
\end{eqnarray}
where $e_Q$ is the electric charge of the heavy quark, $\alpha_{em}$ is the electromagnetic coupling constant and $\beta = 4 m_Q^2/W_{\gamma g}^2$.
The distribution $g^D (x,\mu^2)$ can be obtained by the convolution of the Pomeron flux $f_{\pom} (x_{\pom})$ and the gluon distribution in the Pomeron $g_{\pom} (\beta, \mu^2)$:
\begin{eqnarray}
g^D (x,\mu^2) = \int dx_{\pom} d\beta \delta(x - x_{\pom}\beta) g_{\pom}(\beta,\mu^2) f_{\pom}(x_{\pom}) = \int_x^1 \frac{dx_{\pom}}{x_{\pom}}  f_{\pom}(x_{\pom})g_{\pom}(\frac{x}{x_{\pom}},\mu^2) \,\,\,,
\end{eqnarray} 
where $\mu^2 = 4 m_Q^2$, $x_{\pom}$ is a fraction of the hadron´s momentum transferred into the diffractive system and $\beta$ is a fraction of the Pomeron momentum carried by the gluon participating in the diffractive scattering. The  Pomeron flux $f_{\pom}(x_{\pom})$ enters in the form integrated over the four momentum transfer
\begin{eqnarray}
f_{\pom}(x_{\pom}) = \int_{t_{min}}^{t_{max}} dt f_{\pom}(x_{\pom}, t)\,\,,
\end{eqnarray}
where $t_{min}$ and $t_{max}$ are kinematic boundaries. 
In our calculations we will consider the  Pomeron flux factor and gluon distribution in the Pomeron obtained by the H1 Collaboration in its analyses of the diffractive structure function at HERA
 \cite{H1diff}, which assume a flux factor motivated by Regge theory with  $x_{\pom}$ dependence given by,
\begin{eqnarray}
f_{\pom/p}(x_{\pom}, t) = A_{\pom} \cdot
\frac{e^{B_{\pom} t}}{x_{\pom}^{2\alpha_{\pom} (t)-1}} \ .
\label{eq:fluxfac}
\end{eqnarray}
The Pomeron trajectory is assumed to be linear, $\alpha_{\pom} (t)= \alpha_{\pom} (0) + \alpha_{\pom}^\prime t$, and the parameters $B_{\pom}$ and $\alpha_{\pom}^\prime$ and their uncertainties are obtained from fits to H1 FPS data \cite{H1FPS}.
Moreover, the Pomeron structure function has been modelled in terms of a light flavour singlet distribution $\Sigma(z)$, consisting of $u$, $d$ and $s$ quarks and anti-quarks with $u=d=s=\bar{u}=\bar{d}=\bar{s}$, and a gluon distribution $g(z)$, with $z$ being the longitudinal momentum fraction of the parton entering the hard sub-process with respect to the diffractive exchange, such that $z=\beta$ for the lowest order quark-parton model process, whereas $0<\beta<z$ for higher order processes.

\section{Results}
\label{sec:results}

In what follows, we will compute the rapidity distribution and total cross sections for the diffractive photoproduction of heavy quarks  in proton-proton  and proton-nucleus at high energies. The resolved Pomeron model shortly reviewed in the previous section serve as input for the numerical calculations using Eq. (\ref{sighh}) for  the energies of the  current and future $pp$ and $pA$ accelerators.  Namely, one considers $\sqrt{S_{NN}}=0.2 \, (1.96) $ TeV for $pp (\bar{p})$ collisions at RHIC (TEVATRON). At LHC we assume $\sqrt{S_{NN}}= 7$ and 14 TeV for $pp$ collisions and $\sqrt{S_{NN}}=8.8$ TeV for $pPb$ collisions.
Moreover, we assume in our calculations $m_c = 1.5$ GeV and $m_b = 4.5$ GeV.

The distribution on rapidity $Y$ of the produced open heavy quark state can be directly computed from Eq. (\ref{sighh}), by using its  relation with the photon energy $\omega$, i.e. $Y\propto \ln \, (\omega/m_Q)$.  Explicitly, the rapidity distribution is written down as,
\begin{eqnarray}
\frac{d\sigma \,\left[h_1+h_2 \rightarrow h_1 \otimes Q\overline{Q} X \otimes h_2) \right]}{dY} = \omega \, \frac{dN_{\gamma/h_1} (\omega )}{d\omega }\,\sigma_{\gamma h_2 \rightarrow Q\overline{Q} h_2}\,\left(\omega \right) + \omega \, \frac{dN_{\gamma/h_2} (\omega )}{d\omega }\,\sigma_{\gamma h_1 \rightarrow Q\overline{Q} h_1}\,\left(\omega \right)\,,
\end{eqnarray}
where $X$ is a hadronic final state resulting of the Pomeron fragmentation.

\begin{figure}
\includegraphics[scale=0.2]{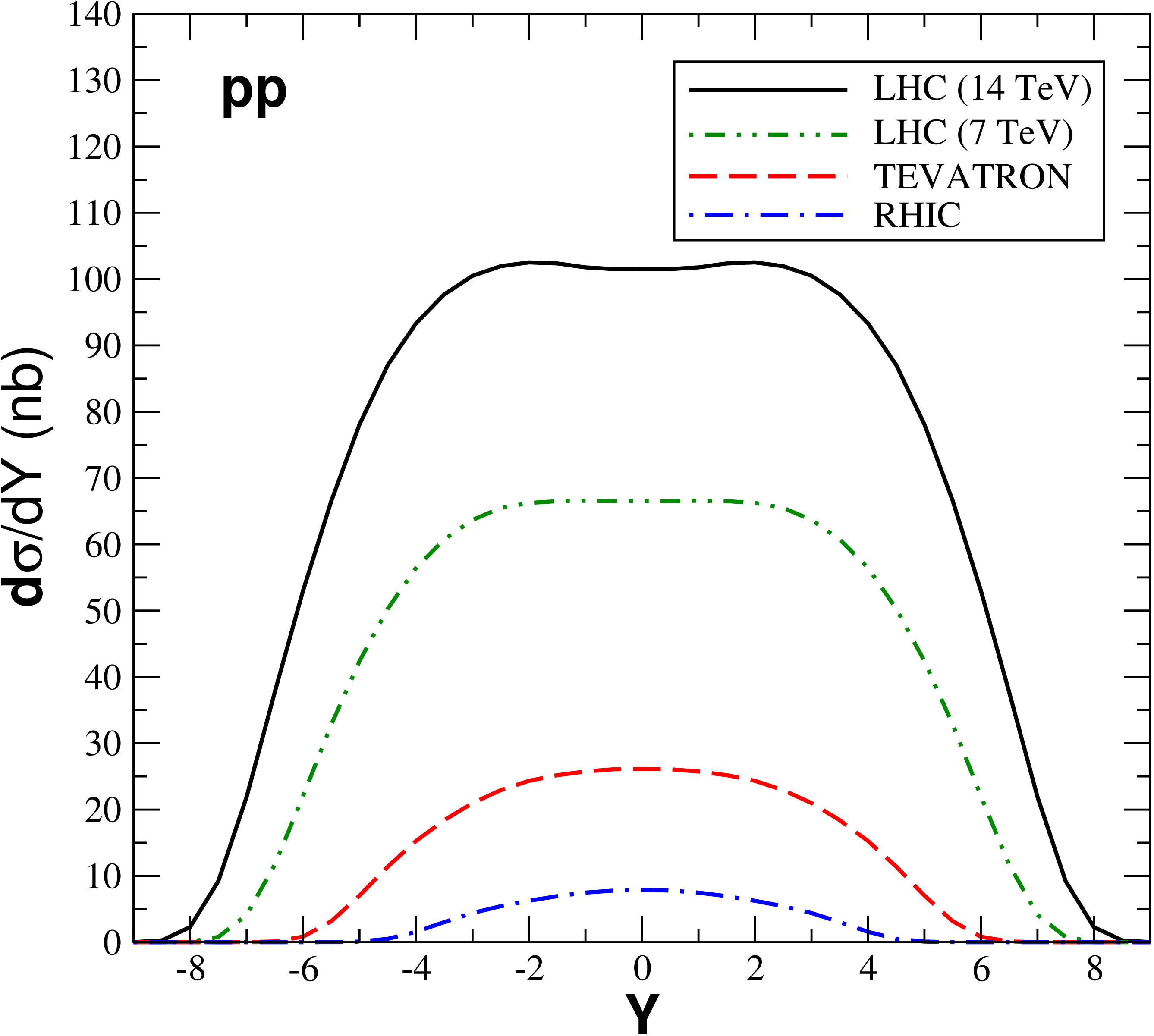}
\caption{Energy dependence of the rapidity distribution for the diffractive charm photoproduction in $pp (\bar{p})$ collisions considering the resolved Pomeron model.}
\label{fig1}
\end{figure}

\begin{table}
\begin{center}
\begin{tabular} {||c|c|c|c|c||}
\hline
\hline
& RHIC & TEVATRON & LHC (7 TeV) & LHC (14 TeV)  \\
\hline
\hline
$\sigma (pp \rightarrow p \otimes c\bar{c} X \otimes p$  & 47 nb   & 214 nb  & 709 nb & 1208 nb \\
\hline
\hline
\end{tabular}
\end{center}
\caption{The total cross section  for the diffractive charm photoproduction at the RHIC, TEVATRON and LHC energies considering the resolved Pomeron model.}
\label{tab1}
\end{table}

\begin{figure}
\begin{tabular}{cc}
\includegraphics[scale=0.15]{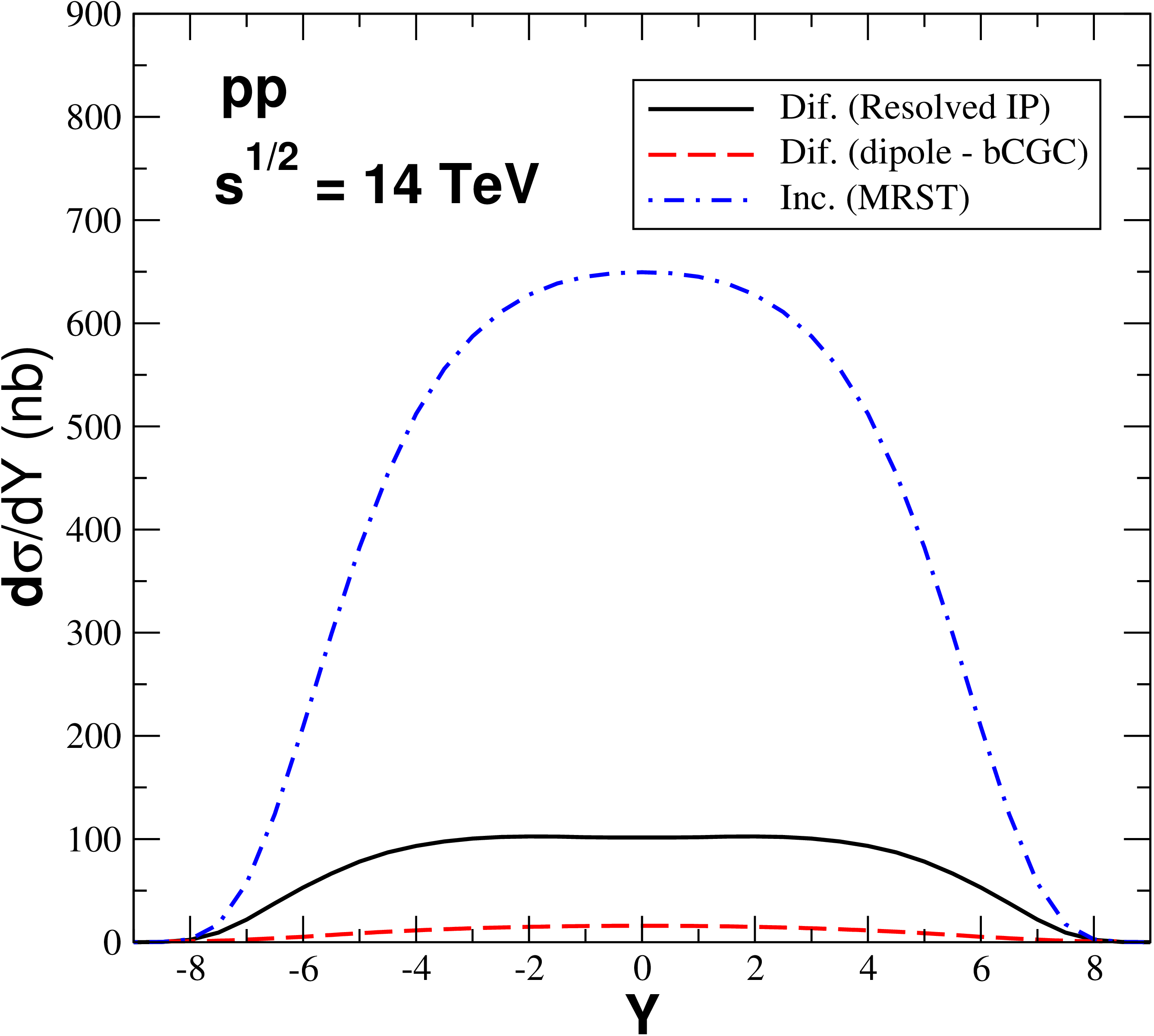} & \includegraphics[scale=0.15]{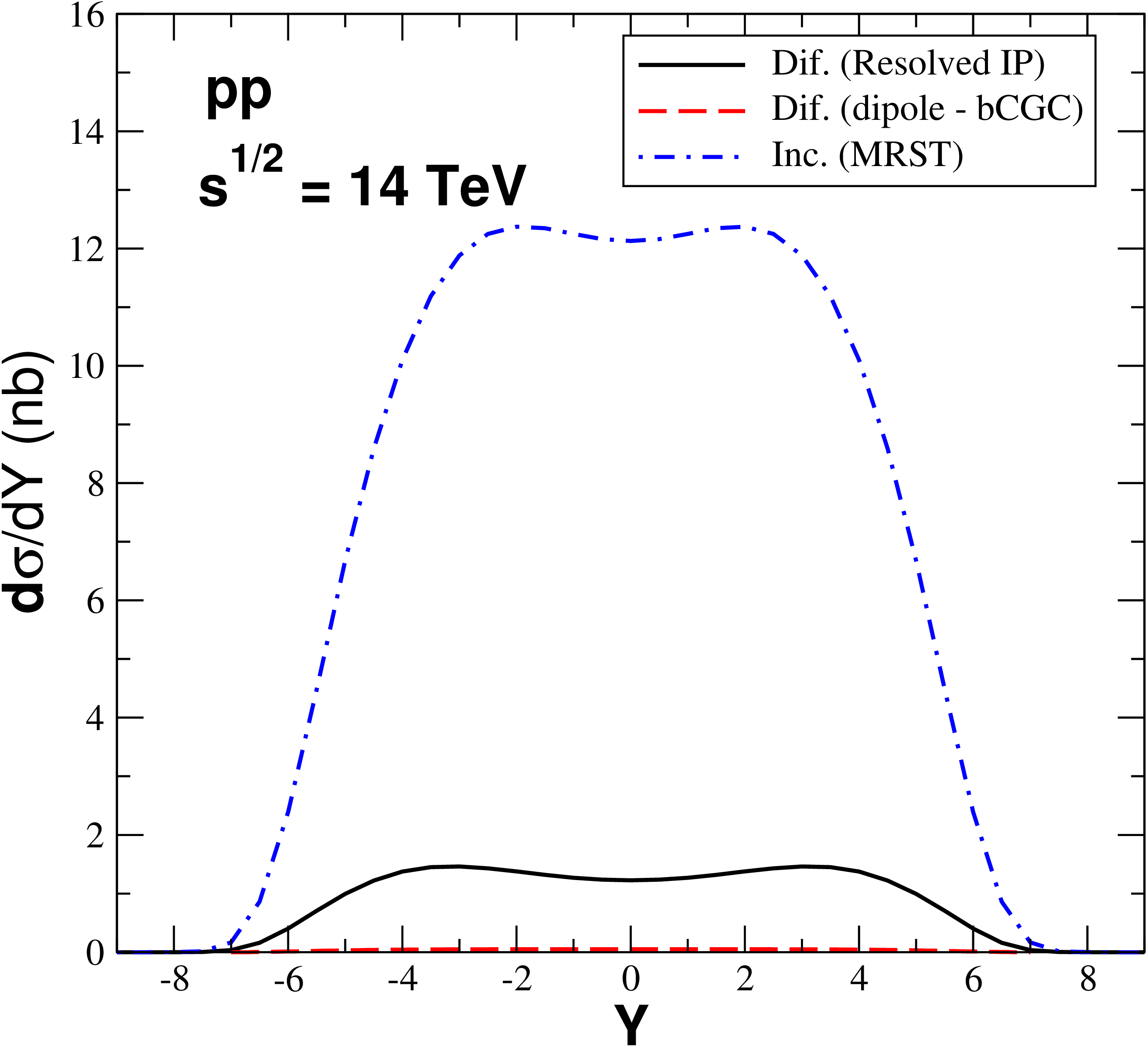}
\end{tabular}
\caption{Rapidity distribution for the inclusive and diffractive charm (left panel) and bottom (right panel) photoproduction  in $pp$ collisions at $\sqrt{s} = 14$ TeV. }
\label{fig2}
\end{figure}

In Fig. \ref{fig1} we present our predictions for energy dependence of the rapidity distribution for diffractive charm photoproduction. As expected from the behaviour of the diffractive gluon distribution at small-$x$, we have that the rapidity distribution increases with the energy. In Table \ref{tab1} we present our predictions for the total cross section. In particular, we predict a value of 709 (1208) nb for $pp$ collisions at 7 (14) TeV. Assuming the  design luminosity     
${\cal L} = 10^7$ mb$^{-1}$s$^{-1}$ the corresponding event rates will be $\approx$ 7000 (12000) events/second.
In order to compare with previous estimates, in Fig. \ref{fig2} we present our predictions for the diffractive charm and bottom photoproduction in $pp$ collisions at $\sqrt{s} = 14$ TeV (denoted resolved Pomeron) and those obtained in \cite{vicmag_ane} using the dipole approach and the bCGC model \cite{bcgc} for the dipole-proton cross section (denoted dipole-bCGC). For comparison we also present the prediction for the inclusive heavy quark photoproduction obtained using as input in the calculation the proton gluon distribution as given by the MRST parametrization \cite{mrst}. Moreover, in Table \ref{tab2} we present our predictions for the total cross section. In the inclusive case,  the values for charm and bottom production are either large, going from some units of nanobarns at bottom  to  microbarns at charm. Therefore, these reactions can have high rates at the  LHC kinematical regime.
On the other hand, the cross sections for diffractive production considering the resolved Pomeron model are a factor $\approx$  5 smaller than the inclusive case. In comparison with the predictions of the dipole model, our predictions are a factor 
$ \gtrsim 7$ larger. It is important to emphasize that although the smaller values of the cross sections for diffractive production  in comparison to the inclusive case, due the clear experimental signature of the diffractive  processes, two rapidity gaps and two intact hadrons in the final state, the experimental analyses is expected to be feasible. 

\begin{table}
\begin{center}
\begin{tabular} {||c|c|c|c|c||}
\hline
\hline
& & Inclusive & Diffractive  & Diffractive   \\
& & &  Dipole model & Resolved Pomeron \\
\hline
\hline
{\bf Charm} & $pp$   & 6697 nb   & 161 nb  & 1208 nb \\
\hline
& $pPb$  & 5203  $\mu b$ & 145  $\mu b$ & 694 $\mu b$ \\
\hline
\hline
{\bf Bottom} & $pp$   & 123 nb   & 0.52 nb  & 15 nb \\
\hline
 & $pPb$  & 55 $\mu b$ & 0.2 $\mu b$ & 4.5 $\mu b$ \\
\hline
\hline
\end{tabular}
\end{center}
\caption{Comparison between the total cross sections for the inclusive and diffractive charm and bottom photoproduction in $pp$ and $pPb$ collisions at LHC.}
\label{tab2}
\end{table}

The diffractive heavy quark photoproduction can also be studied  in proton - nucleus collisions at LHC, which are expected to occur in the next year. In this case the cross sections are enhanced by a factor $Z^2$ present in the photon flux of a nuclei. 
In Fig. \ref{fig3} we present our predictions for the inclusive and diffractive heavy quark photoproduction considering $pPb$ collisions at $\sqrt{s} = 8.8$ TeV. In this case the rapidity distribution is asymmetric due to the dominance of the ion as the photon emitter.  The large difference between the predictions from different models observed in $pp$ collisions also is present in the $pA$ case. In particular, as observed in Table \ref{tab2}, we predict large values for the total cross section, which becomes the experimental test of our results feasible. It is important to emphasize that the predictions of the dipole model for $pA$ collisions are presented here for the first time. Finally, we would like to point out that the diffractive photoproduction should also occur in $AA$ collisions, but in order to estimate the cross section using the resolved Pomeron model, we would need to specify the  diffractive parton distribution of the Pomeron and its flux in a nuclei, which are currently not constrained by experimental data. We postpone the analysis of this case for a future publication.

{Lets discuss the experimental separation of the diffractive photoproduction of heavy quarks and compare our predictions with those obtained considering Pomeron - Pomeron interactions . 
In comparison to the  inclusive heavy quark hadroproduction (See e.g. \cite{rauf}), which is characterized by the process $p + p \rightarrow X + Q\bar{Q} + Y$, with both proton producing hadronic final states, the 
  photoproduction cross sections are a factor $\approx 10^3$ smaller. However, as 
diffractive photoproduction is an exclusive reaction, $p+p \rightarrow p \otimes Q\bar{Q} \otimes p$,  the separation of the signal from hadronic background would be relatively clear. Namely, the characteristic features in photoproduction at coherent collisions are low $p_T$ heavy quark spectra and a double rapidity gap pattern. Moreover, the detection (Roman pots) of the scattered protons can be an additional useful feature. In hadroproduction, the spectra on transverse momentum of heavy quarks are often peaked around heavy quark pair mass, $p_T\approx m_{Q\bar{Q}}$.  Therefore, we expect that a cut in the transverse momentum of the pair, for instance $p_T< 1$ GeV, could eliminate most part of the  contribution associated to the hadroproduction of heavy quarks (See e.g Refs.  \cite{vicmag_inc,vicmag_hqdif}). Furthermore, the rapidity cut would enter as an auxiliary separation mechanism. This procedure is specially powerful, since there will be a rapidity gap on  both sides of the produced heavy quarks in the  diffractive case.  Moreover, in comparison with the hadroproduction of heavy quarks, the event multiplicity for photoproduction interactions is lower, which implies that it may be used as a separation factor between these processes.

\begin{figure}
\begin{tabular}{cc}
\includegraphics[scale=0.15]{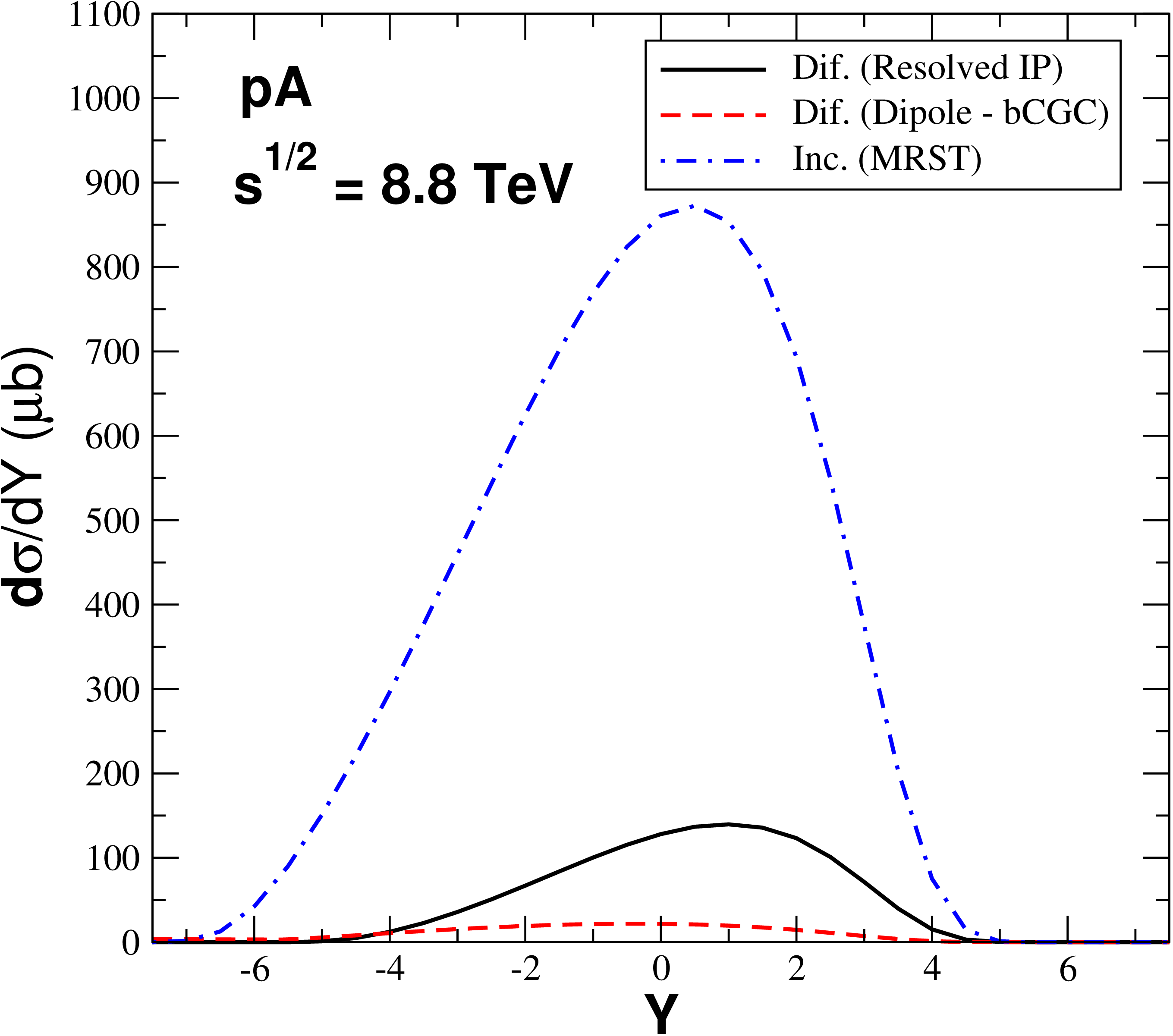} & \includegraphics[scale=0.15]{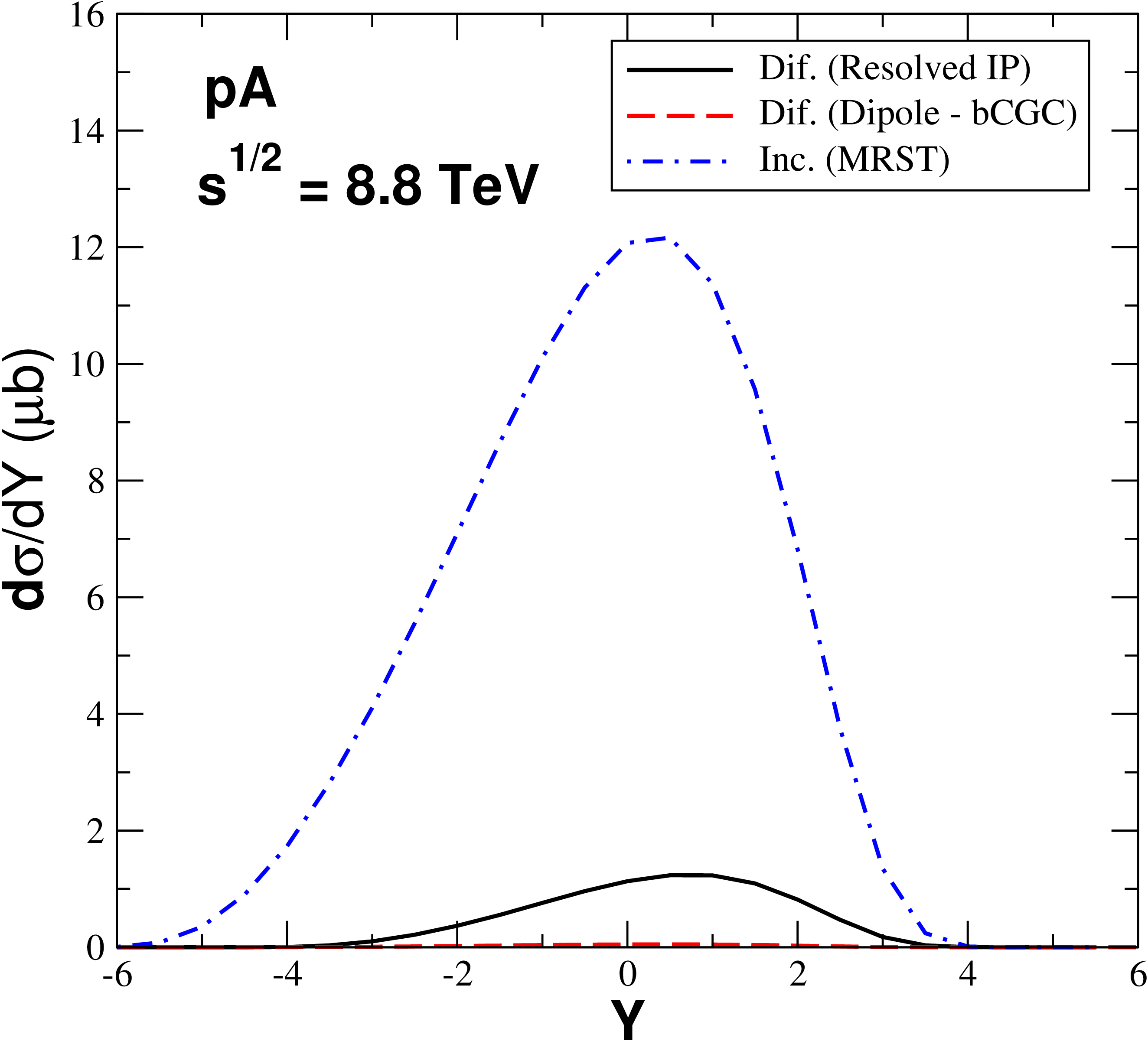}
\end{tabular}
\caption{Rapidity distribution for the inclusive and diffractive charm (left panel) and bottom (right panel) photoproduction  in $pPb$ collisions at $\sqrt{s} = 8.8$ TeV. }
\label{fig3}
\end{figure}

In the case of diffractive photoproduction of heavy quarks  
we expect the presence of two rapidity gaps in the final state, similarly to two - photon or Pomeron - Pomeron interactions. Consequently, it is important to determine the magnitude of this cross section in order to estimate the background for these other channels.  Lets start considering proton - proton collisions. In comparison to production of heavy quark in two - photon interactions \cite{vicmag_hq}, our predictions are a factor  $\approx 10^4$ larger, which implies that this contribution can be disregarded.  In contrast, in comparison to 
the inclusive double diffractive heavy quark production studied in Ref. \cite{mag_hqdif}, which predict 
$\sigma_{c\bar{c}}^{\mathrm{\pom \pom}}\simeq 18$ $\mu$b and $\sigma_{b\bar{b}}^{\mathrm{\pom \pom}}\simeq 0.5$ $\mu$b for LHC energies, our predictions are a factor $\gtrsim 10$ smaller. 
In comparison to the exclusive double diffractive production of charm quarks presented in \cite{antoni_plb} at Tevatron energy, it is a factor $\approx 30$ smaller. It is important to emphasize that these predictions are strongly dependent on the value used for the gap survival factor, while our results should not be modified by soft absorption corrections. 

In the case of proton - nucleus collisions, the magnitude of the heavy quark production considering  Pomeron - Pomeron interactions still is an open question. In particular, the value of the gap survival factor in nuclear collisions is a subject that deserves  a detailed study. Preliminary studies, as in Ref. \cite{miller}, indicate that  this factor is very small ($ \approx 8 \times 10^{-4}$), which implies that  $\gamma \pom$ interactions can dominate the heavy quark production in $pA$ collisions. 

Finally, although our predictions are smaller than those obtained considering double pomeron interactions, it is expected that emerging hadrons from double Pomeron processes have a much larger transverse momentum than those resulting from diffractive photoproduction processes. Consequently, in principle it is possible to introduce a selection criteria  to separate these two processes. However, this subject deserves more detailed studies.}

\section{Conclusions}
\label{sec:conc}

In summary, we have computed the cross sections for  diffractive  photoproduction of heavy quarks in $pp$ and $pA$  collisions at  LHC energies. This has been performed using the resolved Pomeron model based on the diffractive factorization  formalism and that the Pomeron has a partonic structure, which describe quite well the HERA experimental data for the diffractive structure function.   The obtained values are shown to be sizeable  at  LHC. The feasibility of detection of these reactions is encouraging, since their experimental signature should be suitably clear.  Furthermore, considering the large difference between the predictions of the distinct models, they enable to constrain the underlying model for the Pomeron, which is fundamental to predict the observables which will be measured in  hadron-hadron collisions at LHC.

\section*{Acknowledgments}
This work was supported by CNPq, CAPES and FAPERGS, Brazil.

\end{document}